\documentclass[journal=jpcbfk,manuscript=article]{achemso}

\usepackage{bm}
\usepackage{amsmath}
\usepackage{amssymb}
\usepackage{amsfonts}
\usepackage{amsthm}
\usepackage{amscd}
\usepackage{color}
\usepackage{graphicx}
\usepackage[normalem]{ulem}

\def\be{\begin{equation}}
\def\ee{\end{equation}}
\def\eqref#1{(\ref{#1})}

\def\bra#1{\mathinner{\langle{#1}|}}
\def\ket#1{\mathinner{|{#1}\rangle}}
\def\braket#1{\mathinner{\langle{#1}\rangle}}

{\catcode`\|=\active
  \gdef\set#1{\mathinner{\lbrace\,{\mathcode`\|"8000\let|\midvert #1}\,\rbrace}}
  \gdef\Set#1{\left\{\:{\mathcode`\|"8000\let|\SetVert #1}\:\right\}}}
\def\midvert{\egroup\mid\bgroup}
\def\SetVert{\egroup\;\mid@vertical\;\bgroup}

\newcommand{\proj}[2]{\ket{#1}\!\bra{#2}}
\newcommand{\onlinecite}[1]{\cite{#1}}

\author{Piet Schijven}
\email{petrus.schijven@physik.uni-freiburg.de}
\author{Lothar M\"uhlbacher}
\email{lothar.muehlbacher@physik.uni-freiburg.de}
\author{Oliver M\"ulken}
\email{muelken@physik.uni-freiburg.de}
\affiliation{Physikalisches Institut, Universit\"at Freiburg, Hermann-Herder-Strasse 3, 79104 Freiburg, Germany}
\date{\today}

\title{Energy transfer properties and absorption spectra of the FMO
complex: from exact PIMC calculations to TCL
master equations}

\begin{document}

\begin{abstract}
  We investigate the excitonic energy transfer (EET) in the
  Fenna-Matthews-Olsen complex and obtain the linear absorption spectrum (at
  300 K) by a
  phenomenological time-convolutionless (TCL) master equation which is validated
  by utilizing Path Integral Monte Carlo (PIMC) simulations. By applying Marcus'
  theory for choosing the proper Lindblad
  operators for the long-time incoherent hopping process and using local
  non-Markovian dephasing rates, our model shows very good agreement with the
  PIMC results for EET. It also correctly reproduces the linear absorption
  spectrum that is found in experiment, without using any fitting parameters.
\end{abstract}

\section{Introduction}

Since the seminal experiments on the Fenna-Matthews-Olsen (FMO) complex
performed by the Fleming group in 2007 \cite{Engel2007} and subsequently
confirmed by the Engel group in 2010 \cite{Panitchayangkoon2010}, many
theoretical models have been proposed in order to properly describe the
observed long-lasting oscillations in the 2D spectroscopic data. While many
methods are based on quantum master equations of some sort or the other, like
Lindblad or Redfield equations \cite{Wu2010, Palmieri2009, Mohseni2008,
  Rebentrost2009, Caruso2009, Chin2010}, only recently a more detailed description of
  dissipative effects has been attempted in the form of Path Integral Monte Carlo (PIMC) calculations based
  on results from atomistic modeling combining molecular-dynamics (MD)
  simulations with electronic structure calculations
  \cite{Muehlbacher2012}. In contrast to a phenomenological modeling of the
``environment'' (protein scaffold, water molecules, etc.), the latter approach
utilizes BChl-resolved spectral densities which
can be directly incorporated in the PIMC calculations to obtain an exact
account of the full exciton dynamics.

Most quantum master equation approaches use special analytical forms of
spectral densities, such as Ohmic or Lorentzian forms. Although this allows to
obtain solutions for the equations, it remains to be a phenomenological
ansatz. So far the results that have been obtained by hierarchical
time-nonlocal master equations with a Lorentzian spectral density show
reasonable agreement with the experimental results of both the absorption and
2D spectra at 77 K \cite{Chen2011, Hein2012}. 

In contrast, numerically exact methods like PIMC
simulations~\cite{Muhlbacher2004} or the QUAPI
method~\cite{Makri1995a,Makri1995b, Thorwart2004} have proven to be capable to
produce the exact quantum dynamics of excitonic energy transfer over the
experimentally relevant timescales. However, this comes at the price of rather
large computational costs.  Here, we combine the respective strengths of
numerically exact and approximative methods while overcoming their respective
weaknesses: we use PIMC results for the exact quantum dynamics of the excitonic
population dynamics over intermediate timescales
(i.e.~600fs) which still allow for a fast production of the respective
results, yet are sufficient to allow for a comparison to a
time-convolutionless (TCL)
master equation based on a Lindblad approach; we further corroborate our
  results by comparison with PIMC data for up to 1.5 ps. For a model dimer
systems the authors have already successfully demonstrated such a concept
\cite{Mulken2010a}.  Here, our ansatz captures the long-time behavior by
relating to Marcus' theory of electron transport \cite{Marcus1956, Nitzan2006}
because we assume that the
long-time behavior is governed by a classical hopping process between the
individual sites \cite{Nitzan2006, Ishizaki2011}, see below. 
For the
short-time dephasing behavior we introduce non-Markovian dephasing rates
\cite{BreuerOpenQS}.  Since, in principle, we then obtain results for
arbitrarily long times, we have an efficient yet very accurate way to calculate
arbitrary transfer properties. Furthermore, we use the master equation to
calculate the absorption spectrum of the FMO complex (at 300 K), an observable
which can straightforwardly be obtained experimentally
\cite{Freiberg1997}. This opens the possibility to estimate the validity of the
underlying microscopic Hamiltonian as well as its respective parametrization
based on recent results from mixed quantum-classical
simulations~\cite{Olbrich2011a} by getting into direct contact with
experimental data.

\section{Energy transfer on the FMO complex}
\textbf{Microscopic description} The dynamics of single excitations on the FMO
complex is often described by a tight-binding Hamiltonian with 7 localized
sites, corresponding to the 7 bacteriochlorophylls (BChls) of the FMO monomer
\cite{Fenna1975, Brixner2005}.  The influence of the protein scaffold and
solvent on the excitonic dynamics is treated, in the spirit of the
Caldeira-Leggett model \cite{Caldeira1983}, as a collection of harmonic modes
that are linearly coupled to each BChl.  Previous studies showed no significant
correlations between the bath induced energy fluctuations at different sites
\cite{Olbrich2011b, Shim2012}, so we assume that each BChl is coupled to its
own individual environment. The full Hamiltonian of the system can now be
written as: \be \label{full Hamiltonian} H = H_S + H_B + H_{SB}, \ee with
\begin{eqnarray}
 H_S &=& \sum_n \epsilon_n \proj{n}{n} + \sum_{m\neq n} J_{mn} \proj{m}{n} \,, \\
 H_B &=&  \sum_{n,\kappa} \left(\frac{P_{n\kappa}^2}{2m_{n\kappa}} + \frac{1}{2}m_{n\kappa}\omega_{n\kappa}^2 X_{n\kappa}^2\right) \,, \\
 H_{SB} &=&  \sum_{n} \ket{n}\bra{n} \left( c_{n\kappa} X_{n\kappa} + \Lambda_n^{\mathrm{(cl)}} \right)\, .
\end{eqnarray}
The state $\ket{n}$ corresponds to the single-excitation state of site $n$, the
parameter $\epsilon_n$ denotes the energy gap between ground and excited state
of site $n$, and $J_{mn}$ describes the excitonic coupling between sites $m$
and $n$.  Furthermore, $X_{n\kappa}$, $P_{n\kappa}$, $m_{n\kappa}$ and
$\omega_{n\kappa}$ denote the position, momentum, mass and frequency of the
bath oscillators, respectively. In the interaction Hamiltonian $H_{SB}$, the
constants $c_{n\kappa}$ (in units of eV/m) denote the coupling strength between site $n$ and the
bath modes. We have included the classical reorganization energies
$\Lambda_n^{\mathrm{(cl)}}$ as a counter-term in $H_{SB}$ to prevent further
renormalization of the site energies by the environment \cite{Caldeira1983,
  Caldeirra1983Erratum, BreuerOpenQS}.  This quantity is defined as: 
\be
\Lambda_n^{\mathrm{(cl)}} = \frac{\hbar}{\pi} \int_0^\infty d\omega \frac{J_n(\omega)}{\omega}, 
\ee 
where $J_n(\omega)$ (in units of 1/s) is the spectral density of
the bath that is coupled to site $n$. In terms of the system parameters, it is
given by: 
\be 
J_n(\omega) = \frac{\pi}{\hbar}\sum_\kappa \frac{c_{n\kappa}^2}{2 m_{n\kappa}\omega_{n\kappa}} \delta( \omega - \omega_{n\kappa} ) .  
\ee 
The precise numerical values of the different parameters entering in the
expressions above were obtained from combined quantum-classical simulations for
the full FMO complex including the solvent \cite{Olbrich2011, Olbrich2011a}.

\subsection{Effective master equation approach} 
We use now use the microscopic description of the FMO complex to set up a
phenomenological second order time-local quantum master equation. In doing so,
we are able to reproduce the dynamics
obtained from the PIMC simulations as well as extending it to, in
principle, arbitrary long times. Additionally, our approach
also allows to obtain results for the linear absorption spectrum which are in
close accordance to experimental findings.

The spectral density of the FMO complex \cite{Olbrich2011a} leads to
reorganization energies $\Lambda_n^\text{(cl)}$ of the order of $0.02 - 0.09$
eV, which is comparable to the differences in the site energies $\epsilon_n$,
while the excitonic couplings $J_{mn}$ are of the order of 1 meV. This implies
that we can expect that the protein environment is relatively strongly coupled
to the FMO complex and that it therefore leads to a strong damping for the
population dynamics. This is also reflected by the results of the PIMC
simulations \cite{Muehlbacher2012}.

We now assume that in the long-time limit, after most of the coherences
(i.e.~off-diagonal elements of the reduced density matrix in the site-basis
representation) in the system have decayed, EET can
be described by a classical hopping process between the different sites
(BChl's), that is induced by the protein environment. The transfer rates
$k_{mn}$ have to satisfy detailed balance, ensuring a correct equilibrium
state, and are assumed to follow from Fermi's golden rule. Furthermore, the
rates should also depend on the reorganization energies $\Lambda_n^\text{(cl)}$
and $\Lambda_m^\text{(cl)}$ of the baths that are coupled to the sites $n$ and
$m$, reflecting the differences in the coupling strengths of the protein
environment to each BChl. Unlike F\"orster theory, which assumes incoherent
hopping between the energy eigenstates of $H_s$ \cite{Forster1959}, Marcus's
theory of electron transport satisfies all these properties \cite{Marcus1956,
  Nitzan2006}, leading to transfer rates $k_{mn}$ of the form: 
\be 
k_{mn} =
\sqrt{\frac{\pi \beta}{\hbar^2 \Lambda_{mn}^{\mathrm{(cl)}} }} |J_{mn}|^2
\exp{\left[- \frac{\beta(\epsilon_n - \epsilon_m +
      \Lambda_{mn}^{\mathrm{(cl)}})^2}{4\Lambda_{mn}^{\mathrm{(cl)}}}\right]} ,
\ee 
with $\beta = 1/k_B T$ and $\Lambda_{mn}^{\mathrm{(cl)}} =
\Lambda_m^{\mathrm{(cl)}} + \Lambda_n^{\mathrm{(cl)}}$.

Aside from incoherent transfer between the sites, the environment also induces
a strong dephasing on each site. In the framework of the second order
TCL master equation \cite{BreuerOpenQS}, these dephasing
rates (in units of 1/fs) are given by: 
\be 
\lambda_n(t) = 2\mathrm{Re}\int_0^t ds \,\int_0^\infty d\omega\, J_n(\omega) \left[ \coth{\left(\beta \hbar\omega / 2 \right)}\cos(\omega s) - i \sin(\omega s)\right].  
\ee 
Here, we use the spectral
densities $J_n(\omega)$ which have been obtained by MD simulations in
Ref.~\onlinecite{Olbrich2011a} and numerically calculate the correlation function.

The TCL master equation that describes the excitation dynamics can now be
written as \cite{BreuerOpenQS}: 
\be\label{eq:TCL} 
\frac{d\rho(t)}{dt} \equiv
\mathcal{L}(t)\rho(t) = -\frac{i}{\hbar}[H_s, \rho(t)] + \mathcal{D}(t)\rho(t)
.  
\ee 
Our numerical results (not displayed) show that the Lamb shift term that
usually appears in this equation, only leads to a negligible difference in both
the population dynamics and the linear absorption spectrum (the position of the
peak is shifted by approximately -1 meV).  The dissipator $\mathcal{D}(t)$ is
assumed to take the following Lindblad form, according to the considerations
above \cite{BreuerOpenQS}: 
\be 
\mathcal{D}(t)\rho(t) = \sum_{mn} \gamma_{mn}(t)
\left( L_{mn} \rho L_{mn}^\dagger - \frac{1}{2}\left\{L_{mn}^\dagger L_{mn},
    \rho \right\}\right) .  
\ee 
The Lindblad operators are defined by $L_{mn} =
\proj{m}{n}$ and the rates by $\gamma_{mm}(t) = \lambda_m(t)$ and
$\gamma_{mn}(t) = k_{mn}$ for $m\neq n$. The operators $L_{mm}$ model the
dephasing process, while the operators $L_{mn}$ model the incoherent transfer
between sites $m$ and $n$. This choice of Lindblad operators will lead - in the
long-time limit - to incoherent hopping transfer between the sites, which is 
different from, e.g., Redfield theory, which requires
incoherent transfer between the eigenstates $\ket{\psi}$ of $H_S$, leading to
Lindblad operators of the form $L \sim \proj{\psi_n}{\psi_m}$ \cite{BreuerOpenQS,
  Nitzan2006}.

However, we note that the equilibrium state of our master equation ($\rho_{eq}$) is slightly
different from the one that follows from Marcus theory $\rho_{eq, db}$, which is given by detailed
balance, $\lim_{t\to\infty} \rho_{nn}(t) = (1/Z)\exp(-\beta \epsilon_n)$ and $\lim_{t\to\infty}\rho_{mn}(t) = 0$ for $m\neq n$, 
where $\rho_{mn}(t) = \langle m|\rho(t)|n\rangle$. This can be shown by noting that
$\mathcal{D}(t)\rho_{eq, db} = 0$ but $[H_S, \rho_{eq, db}] \neq 0$. This implies that
$\rho_{eq, db}$ is not a stationary state of our master equation. For the present calculation, the devations
are only of the order of 1\%, so we still expect our approach to give good results.

%
%

\subsection{Path Integral Monte Carlo simulations}

PIMC simulations allow to extract the exact quantum dynamics in the
  presence of a dissipative environment, both for charge
  transport~\cite{Egger1994,Muhlbacher2004,Muhlbacher2005} as well as energy
  transfer~\cite{Muehlbacher2012}. In short, the time evolution of the reduced
  density operator of a dissipative quantum systems is calculated by employing
  the path integral representation for the propagator according to the
  Feynman-Vernon theory \cite{Feynman1963} for factorizing or its extension to
  correlated initial preparations \cite{Grabert1988}. These path integrals are
  then evaluated by a stochastic sampling process based on Markov walks through
  the configuration space of all conceivable quantum paths which
  self-consistently emphasize the physically most relevant ones.  While there
  is no limitation with respect to the choice of system parameters for which
  PIMC simulation are capable of producing numerically exact results, this
  approach is subject to the notorious `dynamical sign problem'
  \cite{Suzuki1993}, which reflects quantum-mechanical interferences between
  different system paths and results in an increase of the computational effort
  necessary to obtain statistically converged results which scales
  exponentially with the timescale over which the dynamics of the system is
  investigated. However, the presence of a dissipative environment
  substantially weakens these interference effects and therefore the sign
  problem. Furthermore, it allows for various efficient optimization schemes
  which lead to a further soothing of this computational bottleneck, thus
  significantly enlarging the accessible timescales
  \cite{Egger2000,Stockburger2002,Muhlbacher2004}.

For the present case, we utilize the PIMC data presented in
  Ref.~\onlinecite{Muehlbacher2012} to demonstrate the reliability of the
  master equation results and extend some of the former to longer
  timescales. To that extend, a factorizing initial preparation has been
  employed, where, resembling the situation prior to the creation of an
  exciton, the bath modes initially are in thermal equilibrium with respect to
  themselves, while the exciton has been modeled to be either initially
  localized on one of the seven BChl sites or in one of the seven excitonic
  eigenstates.

\begin{figure*}
 \begin{center}
  \includegraphics[width=0.8\textwidth]{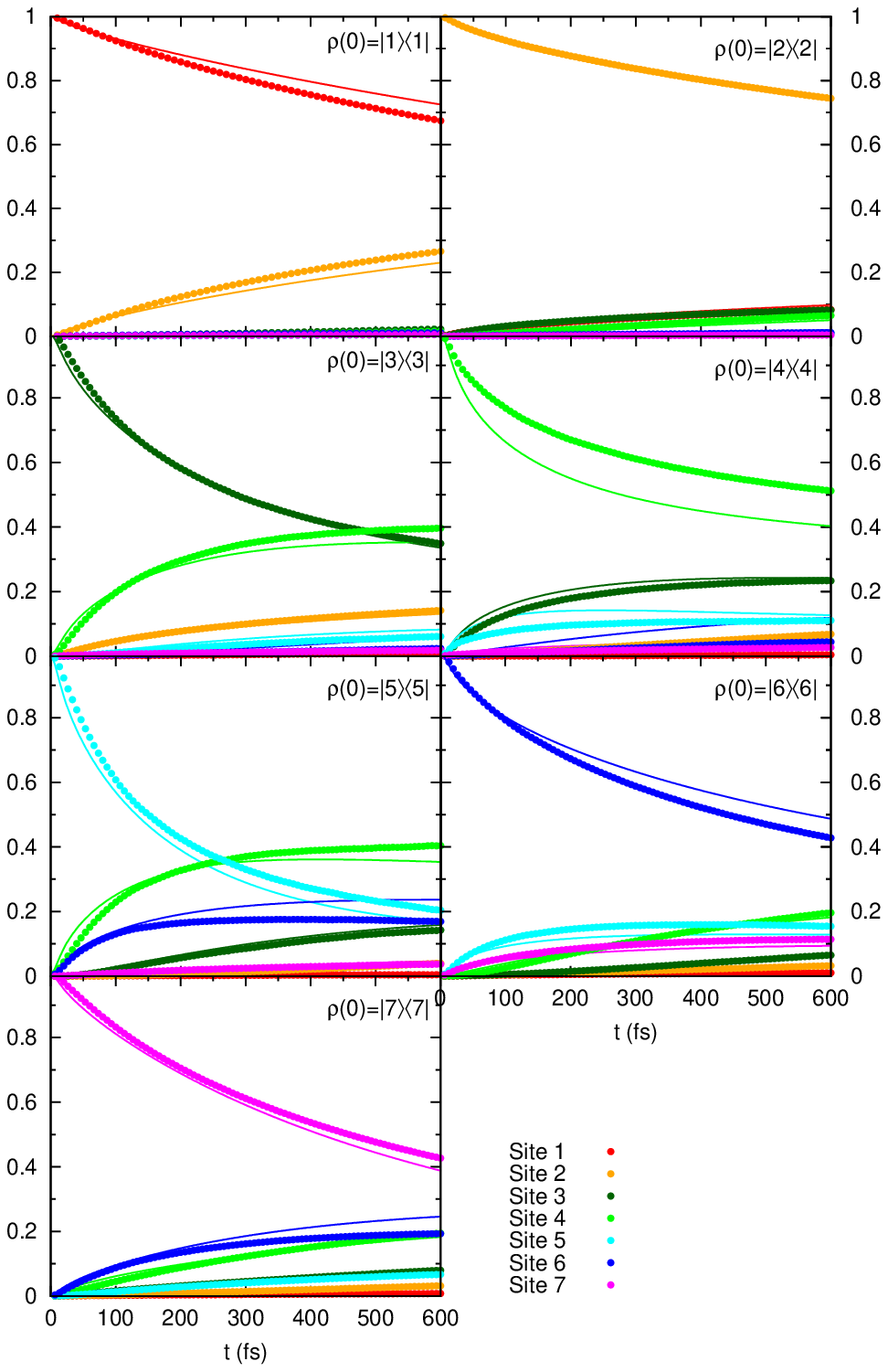}
 \end{center}
 \caption{Comparison of population dynamics of the 7 different sites of the
FMO complex obtained from the numerically exact PIMC results (circles) with the
results from the TCL master equation approach (solid lines) for different localized
initial conditions on the sites $|n\rangle$, $n=1,\dots,7$. Note that the
statistical error of the PIMC calculations is typically smaller than the symbol size. 
Therefore we do not show the error bars.}
\label{fig:localized-prep}
\end{figure*}

\begin{figure}
 \begin{center}
  \includegraphics[width=0.8\textwidth]{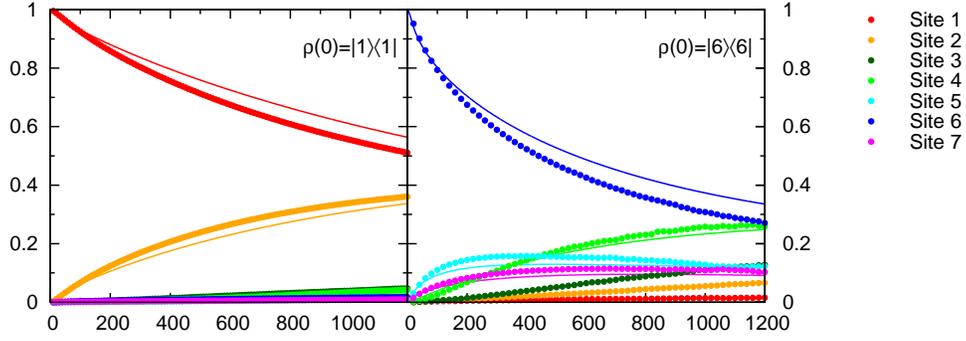}
 \end{center}
 \caption{Same as Fig.~\ref{fig:localized-prep} for localized initial preparations
 in sites 1 and 6, but now extended beyond 1ps}.
\label{fig:localized-prep-lt}
\end{figure}

\begin{figure*}
 \begin{center}
  \includegraphics[width=0.8\textwidth]{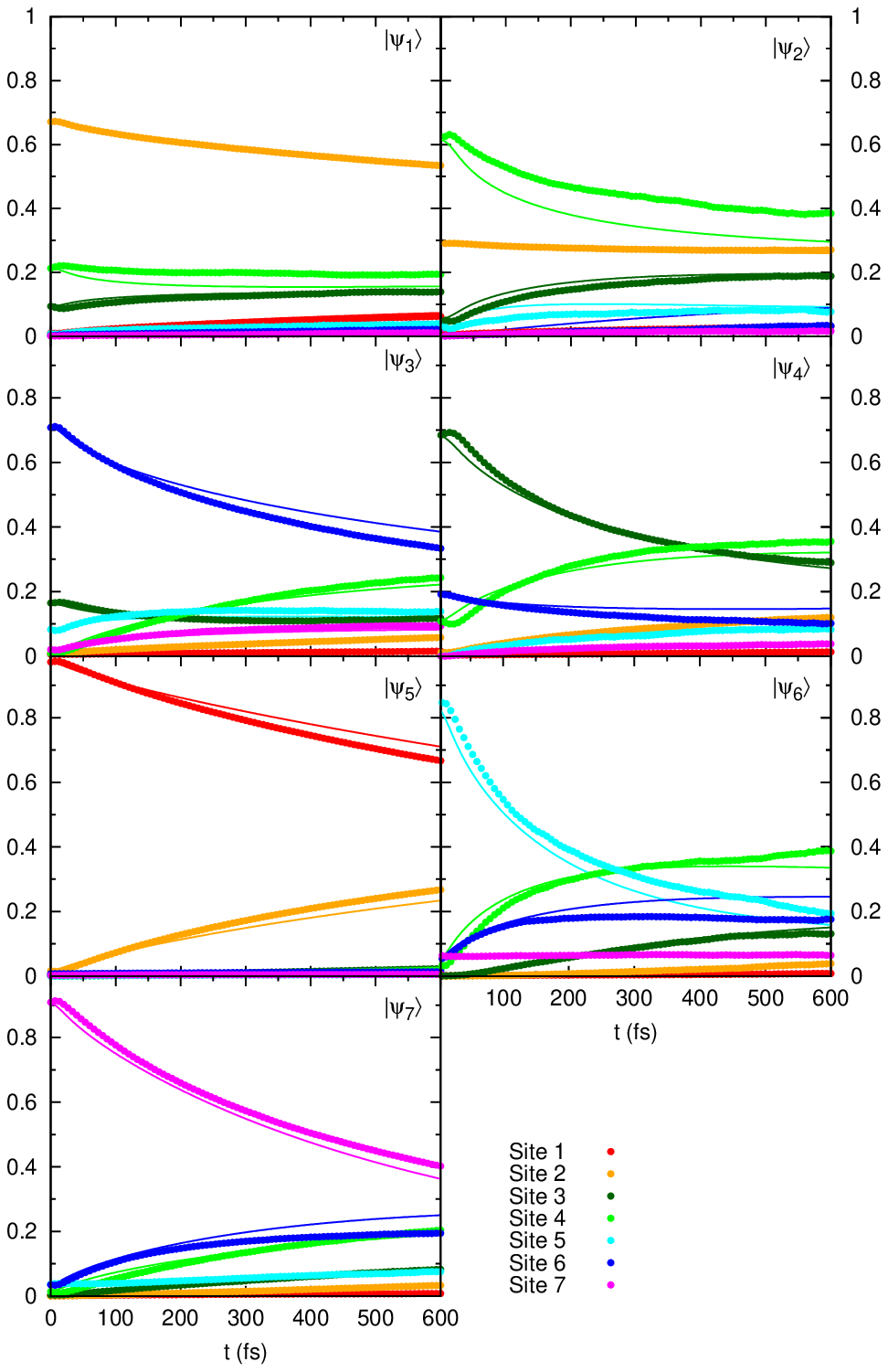}
 \end{center}
 \caption{Same as Fig.~\ref{fig:localized-prep} but taking the eigenstates
$|\Psi_n\rangle$
of $H_S$ as initial conditions.
}
\label{fig:exciton-prep}
\end{figure*}

\subsubsection{Population dynamics}

In Fig. \ref{fig:localized-prep} we show the population dynamics that is
obtained by solving the TCL master equation, Eq. \eqref{eq:TCL}, for initial
conditions corresponding to a localization on
the sites $\ket{n}$, e.g. $\rho(0) = \proj{n}{n}$, in comparison to the
corresponding numerically exact PIMC results.  The dotted curves represents the
latter and the solid lines represent the results from our master equation
approach. Fig. \ref{fig:localized-prep-lt} shows an extension of the results up
to 1200 and 1500 fs for initial preparations in sites 1 and 6,
respectively.

In general we observe good quantitative agreement of our approach with the PIMC
results for all localized initial preparations and over all observed
  timescales. The largest deviations are observed for an initial condition
localized on site 4 for which the bath has the lowest reorganization energy
(0.025 eV). Since our assumption of a classical hopping process at long-times
is based on having a strong coupling to the environment, we would expect that
our approach becomes worse with decreasing reorganization
energy. Also, from Fig. \ref{fig:localized-prep-lt} one observes a good
agreement in the approach to equilibrium, although the decay is slightly
slower than predicted by the PIMC results.

Figure \ref{fig:exciton-prep} corroborates our results. Here, the 
excitonic excitation is initially
in one of the seven eigenstates $|\psi_n\rangle$ of $H_S$.  Again we
find very good agreement with the PIMC results, where once more the strongest
deviations occur for the initial preparation exhibiting the largest population
on site 4.  We note that there is no fitting parameter involved. Introducing a
parameter which interpolates between purely coherent and purely incoherent
transfer, as in \onlinecite{Whitfield2010, Schijven2011,
  Schijven2012}, could lead to a further improvement of the agreement.
Nevertheless, already this rather simple phenomenological model captures most
of the details which are present in the PIMC calculations. Additionally, it
allows for a computationally cheap calculation of the linear absorption
spectrum.

\section{Linear absorption spectrum}
\begin{table*}
\begin{tabular}{c|ccccccc}
  $m$ & 1 & 2  & 3  & 4  & 5 & 6 & 7  \\
  \hline
  ${\mu}_{m, x}$ & 0.0 & -6.10 & -5.27 & 0.0 & -6.39 & 5.16 & 0.0 \\
  ${\mu}_{m, y}$ & 1.86 & 1.08 & -3.04 & 2.49 & 0.0 & 2.98 & -1.14 \\
  ${\mu}_{m, z}$ & 6.07 & 1.66 & -2.10 & 5.85 & -0.45 & 2.29 & 5.85 \\
  $|\vec{\mu}_{m}|^2$ & 40.32 & 41.09 & 41.47 & 40.45 & 41.09 & 40.70 & 35.52  
\end{tabular}
\caption{The numerical values for the x-, y-, and z- component as well as
  the absolute value squared of the transition dipole moments $\vec{\mu}_m =
  ({\mu}_{m, x}, {\mu}_{m, y}, {\mu}_{m, z})$ in units of Debye $[D]$. The $z$ axis is 
  chosen along the $C_3$-symmetry axis of the FMO complex, and the $y$ axis is
  chosen to be parallel to the $N_B-N_D$ axis of BChl 1. }
\label{tab:tdip}
\end{table*} 
Due to the strong dephasing that is induced by the protein environment, the
coherences (i.e., the off-diagonal elements of the reduced density matrix
in the site basis) die out rather fast. Hence, the
dynamics of the populations is mostly insensitive to the exact behavior of the
coherences. However, other observables, such as the linear absorption spectrum,
are very sensitive to the short-time behavior, where the coherences are still
present.  
Being able to access the full dynamics of the reduced
  density matrix for, in principle, arbitrary timescale, now allows us to
  access in particular the linear absorption spectrum for the FMO complex.
We calculate the
linear absorption spectrum and compare it to the absorption
spectrum which was measured in experiment \cite{Freiberg1997} and to the one
which has been computed with mixed quantum-classical simulations by Olbrich
\textit{et~al.} \cite{Olbrich2011a}.  

\begin{figure*}
 \begin{center}
   \includegraphics[width=\textwidth]{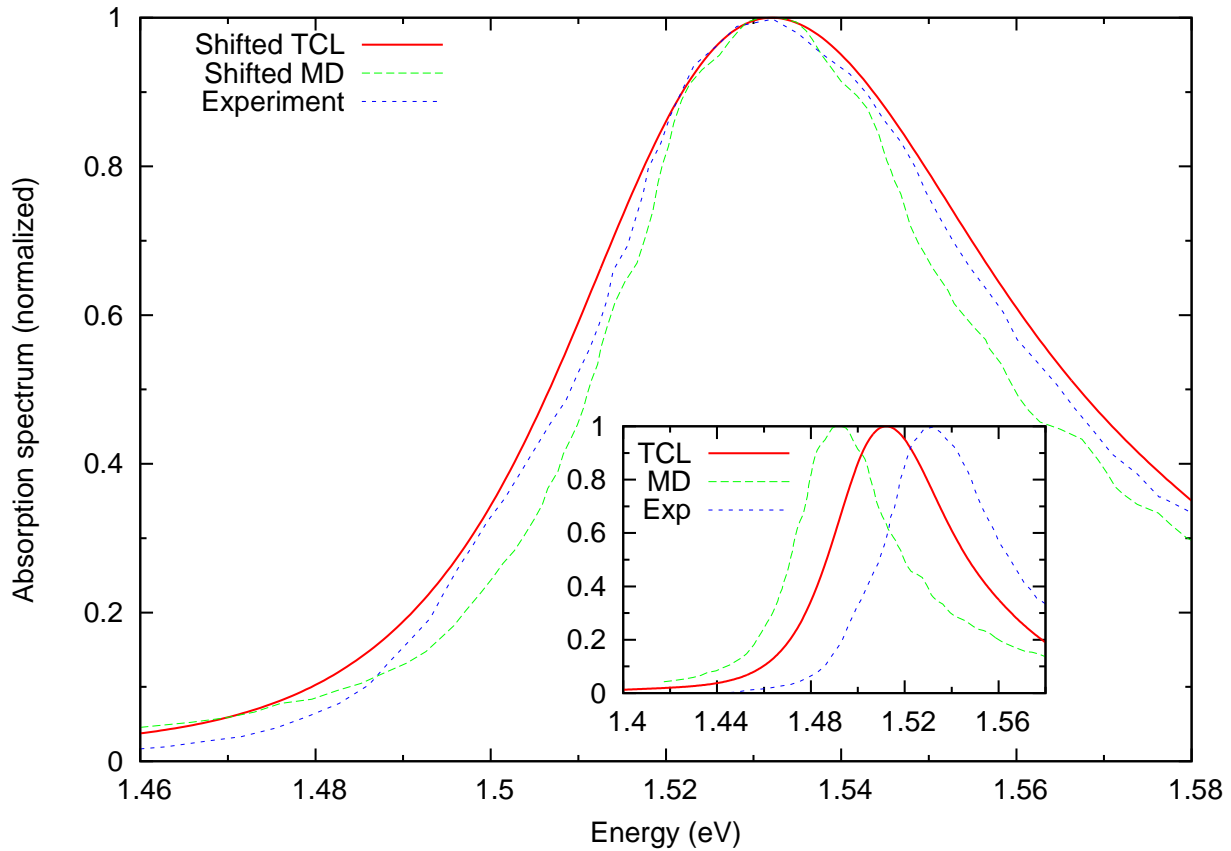}
 \end{center}
 \caption{Comparison of the linear absorption spectra,
   computed with the TCL master equation (solid red),
   mixed quantum-classical calculations~\cite{Olbrich2011a} (dashed green) and obtained by
   experiment~\cite{Freiberg1997} (dotted blue). All three spectra are
overlaid such that the position of the peaks
     are shifted to that
     of the experimental result. The inset shows the spectra at their
original positions.}
\label{fig:absorption}
\end{figure*}

The linear absorption spectrum is given by the Fourier transform of the
two-time correlation function of the transition dipole moment (TDM) operator
$\mu$ \cite{Nitzan2006}: 
\be 
A(\omega) = \mathrm{Re} \int_0^\infty dt\, e^{i\omega t} \braket{ \vec{\mu}(t)\cdot\vec{\mu}(0)}, 
\ee 
where $\vec{\mu}(t) = e^{i Ht/\hbar} \vec{\mu}
e^{-i H t/\hbar}$ is the TDM operator in the Heisenberg picture and $\vec{\mu} =
\sum_m \vec{\mu}_m\left( \proj{m}{0} + \proj{0}{m}\right)$, with $\vec{\mu}_m$
the TDM vector of site $m$. 
The two-time correlation function is evaluated in the
excitonic ground state $W^{0} = \proj{0}{0} \otimes \rho_B$,i.e., with no
excitations present.

To compute this correlation function with our approach, we
use the following expression \cite{BreuerOpenQS}: 
\be\label{eq:autocorr}
\braket{ \vec{\mu}(t)\cdot \vec{\mu}(0)} = \mathrm{tr}_S \left\{ \vec{\mu}(t)\cdot\vec{V}(t) \right\} , 
\ee
where the vector-operator $\vec{V}(t)$ satisfies the TCL master equation 
\be\label{eq:vt}
\frac{d \vec{V}(t)}{dt} = \mathcal{L}(t) \vec{V}(t) , 
\ee 
with the initial condition 
\be
\vec{V}(0) = \vec{\mu} \mathrm{tr}_B\left\{W^{0}\right\} = \vec{\mu} \proj{0}{0} = \sum_m \vec{\mu}_m \proj{m}{0} \,, 
\ee 
where $\mathrm{tr}_S$ and $\mathrm{tr}_B$
denote the trace over the excitonic and environmental degrees of freedom,
respectively.  Due to the time dependence of the dephasing rates that enter
into the generator $\mathcal{L}(t)$, we cannot obtain an analytical solution
for the linear absorption spectrum. Therefore, we solve Eq. \eqref{eq:vt}
numerically and obtain the absorption spectrum with Eq. \eqref{eq:autocorr}.

The numerical values of the TDM vectors $\vec{\mu}_{m}$ are obtained by
combining the data from Refs.~\onlinecite{Olbrich2011a} and $\onlinecite{Blankenship1997}$
together with their relative orientations with respect to the $C_3$ symmetry
axis, taken from Ref.~\onlinecite{Milder2010}. In table \ref{tab:tdip} we provide the
computed values of the $\vec{\mu}_{m}$'s.

In Fig. \eqref{fig:absorption} we show the numerical results for the linear
absorption spectrum and compare it to the experimental results by Freiberg
\textit{et. al.} \cite{Freiberg1997} and computational results obtained by
Olbrich \textit{et. al.}  \cite{Olbrich2011a} utilizing the same
  parametrization of the Hamiltonian, Eq.~(\ref{full Hamiltonian}), but based on
  ensemble-averaged wave packet dynamics~\cite{Parandekar2006}.  
We observe that both the spectra from the latter and from our theoretical
approach are shifted by $-0.04$ eV and $-0.02$ eV, respectively,
compared to the experimental spectrum. When overlaying
those spectra such that the position of the peaks match, we observe almost
perfect agreement for the overall actual line shape, even though our 
method lacks the detailed features of the experiment. 
We note that our approach yields a shift from the experimental findings
which is only half as large as the one presented in Ref.~\onlinecite{Olbrich2011a}.
This indicates that despite the strong
  dissipation, quantum effects still play an important role for the excitonic
  dynamics which are not captured by the ensemble-averaged wave packet
dynamics. 

\section{Summary}

We have presented an approach to calculate the EET as well as the absorption
spectrum of the FMO complex. Our approach is based on a phenomenological TCL
master equation: The dissipator in the master equation is determined by incoherent
hopping rates obtained from Marcus' theory as well as by non-Markovian (pure)
dephasing rates obtained from the bath autocorrelation function. Here, we have
used spectral densities calculated from
  atomistic simulations of Ref.~\onlinecite{Olbrich2011}. 
We have demonstrated
the quantitative reliability of our TCL master equation by a comparison with population dynamics
  obtained from numerically exact PIMC simulations exhibiting an
  excellent accuracy for timescales up to the picosecond
  range, both, for the exciton initially found localized on any
  BChl as well as in an excitonic eigenstate. 
Furthermore, we also found very good agreement with the
experimentally measured absorption spectrum.
Here, the overall line shape of the absorption spectrum is reproduced very
nicely, although the position of the peak is shifted by 0.02 eV as compared to
the experimental result, yet considerably less than for calculations
  based on wave packet dynamics.  Since we have correctly reproduced both the
population dynamics and the dynamics of the coherences, as reflected in the
absorption spectrum, we also expect to be able to reproduce the results for 2D
electronic spectroscopy that were found in experiments at
  ambient conditions, which is subject of current research.

\begin{acknowledgement}
 We gratefully acknowledge support from the Deutsche Forschungsgemeinschaft
 (DFG grant MU2925/1-1 MU 2926/1-1).
 Furthermore, we thank A. Anishchenko, A. Blumen, and L. Lenz for useful discussions.  
\end{acknowledgement}

\bibliographystyle{achemso.bst}
\providecommand*\mcitethebibliography{\thebibliography}
\csname @ifundefined\endcsname{endmcitethebibliography}
  {\let\endmcitethebibliography\endthebibliography}{}

\end{document}